\RequirePackage[2020-02-02]{latexrelease}
\documentclass{iau}

\usepackage{amsmath}
\usepackage{graphicx}

\usepackage{gensymb}
\usepackage{enumerate}

\newcommand{\msun}{\ifmmode{~\rm{M}_{\odot}}\else{M$_{\odot}$}\fi}
\newcommand{\mstar}{\ifmmode{M_{\star}}\else{$M_{\star}$}\fi}

\newcommand{\logm}{\ifmmode{\log(M_{\star}/M_{\odot})}\else{$\log(M_{\star}/M_{\odot})$}\fi}
\newcommand{\mm}{\ifmmode{M_{\star}/M_{\odot}}\else{$M_{\star}/M_{\odot}$}\fi}

\newcommand{\teffe}{$\rm T_{eff}$}

\newcommand{\logge}{$\log{g}$}
\newcommand{\kms}{\mbox{km s$^{-1}~$}} 
\newcommand{\kmse}{\mbox{km s$^{-1}$}}

\newcommand\aap{A\&A}                
\newcommand\aj{AJ}                   
\newcommand\apj{ApJ}                 
\newcommand\araa{ARA\&A}             
\newcommand\mnras{MNRAS}             
\newcommand\nat{Nature}              
\newcommand\pasp{PASP}               

\begin{document}

\lefttitle{Nidever et al.}
\righttitle{The Prevalence of the $\alpha$-bimodality}

\jnlPage{1}{7}
\jnlDoiYr{2021}
\doival{10.1017/xxxxx}

\aopheadtitle{Proceedings IAU Symposium}
\editors{F. Tabatabaei,  B. Barbuy \&  Y. Ting, eds.}

\title{The Prevalence of the $\alpha$-bimodality: \\
First JWST $\alpha$-abundance Results in M31}

\author{David L. Nidever$^{1}$,
Karoline Gilbert$^{2}$,
Erik Tollerud$^{2}$,
Charles Siders$^{1}$, \\
Ivanna Escala$^{3,4}$,
Carlos Allende Prieto$^{5,6}$,
Verne Smith$^{7,8}$,
Katia Cunha$^{8,9,10}$, 
Victor P. Debattista$^{11}$,
Yuan-Sen Ting$^{12,13}$ and
Evan~N.~Kirby$^{14}$
}

\affiliation{$^1${Department of Physics, Montana State University, P.O. Box 173840, Bozeman, MT 59717-3840, USA.
email: {\tt dnidever@montana.edu}} \\
$^2${Space Telescope Science Institute, Baltimore, MD, USA}\\
$^3${Princeton University, 4 Ivy Lane, Princeton, NJ 08544, USA}\\
$^4${The Observatories of the Carnegie Institution for Science, 813 Santa Barbara St., Pasadena, CA 91101, USA}\\
$^5${Instituto de Astrofsica de Canarias, E-38205 La Laguna, Tenerife, Spain}\\
$^6${Departamento de Astrofsica, Universidad de La Laguna (ULL), E-38206 La Laguna, Tenerife, Spain}\\
$^7${NSF’s National Optical-Infrared Astronomy Research Laboratory, 950 North Cherry Avenue, Tucson, AZ 85719, USA}\\
$^{8}$ Institut d'Astrophysique de Paris, UMR 7095 CNRS, Sorbonne Universit\'{e}, 98bis Bd. Arago, 75014 Paris, France
$^{9}$Observat\'{o}rio Nacional, Rua General Jos\'{e} Cristino, 77, 20921-400 S\~{a}o Crist\'{o}v\~{a}o, Rio de Janeiro, RJ, Brazil \\
$^{10}$ Steward Observatory, University of Arizona, 933 North Cherry Avenue, Tucson, AZ, 85721, USA \\
$^{11}${Jeremiah  Horrocks  Institute,  University  of  Central  Lancashire, Preston, PR1 2HE, UK}\\
$^{12}${Research School of Astronomy \& Astrophysics, Australian National University, Cotter Rd., Weston, ACT 2611, Australia}\\
$^{13}${Research School of Computer Science, Australian National University, Acton, ACT 2601, Australia}\\
$^{14}${Department of Physics and Astronomy, University of Notre Dame, 225 Nieuwland Science Hall, Notre Dame, IN 46556, USA}\\
}



\begin{abstract}
We present initial results from our JWST NIRSpec program to study the $\alpha$-abundances in the M31 disk.
The Milky Way has two chemically-defined disks, the low-$\alpha$ and high-$\alpha$ disks, which are closely
related to the thin and thick disks, respectively.
The origin of the two populations and the $\alpha$-bimodality between them is not entirely clear, although there are now several models that can reproduce the observed features.  To help constrain the models and discern the origin, we have undertaken a study of the chemical abundances of the M31 disk using JWST NIRSpec, in order to determine whether stars in M31's disk also show an $\alpha$-abundance bimodality.  Approximately 100 stars were observed in our single NIRSpec field at a projected distance of 18 kpc from the M31 center.  The 1-D extracted spectra have an average signal-to-noise ratio of 85 leading to statistical metallicity precision of 0.016 dex, $\alpha$-abundance precision of 0.012 dex, and a radial velocity precision 8 km s$^{-1}$ (mostly from systematics).
The initial results indicate that, in contrast to the Milky Way, there is {\em no} $\alpha$-bimodality in the M31 disk, and no low-$\alpha$ sequence.  The entire stellar population falls along a single chemical sequence very similar to the MW's high-$\alpha$ component which had a high star formation rate.
While this is somewhat unexpected, the result is not that surprising based on other studies that found the M31 disk
has a larger velocity dispersion than the MW and is dominated by a thick component.  M31 has had a more active accretion and merger history than the MW which might explain the chemical differences.
\end{abstract}


\maketitle
%

\section{Introduction}

\noindent  It has long been known that the Milky Way's disk is composed of a thin and thick component \citep{Yoshii1982,Gilmore1983}.  More recently, it has been determined that the disk is also composed of two chemical populations in the [$\alpha$/Fe]--[Fe/H] plane: the old ($>$8 Gyr), high $\alpha$-abundance population and a younger ($<$8 Gyr), low $\alpha$-abundance population \citep[e.g.,][]{Fuhrmann1998,Reddy2006}.  There exists a valley or ``gap'' between these two populations at intermediate metallicities ([Fe/H]$\approx$$-$1) giving rise to an $\alpha$-bimodality which is clearly seen in the \citet{Fuhrmann2011} volume-complete sample of solar neighborhood stars.
The SDSS/APOGEE survey \citep{Majewski2017} found that this chemical feature is widespread across the MW disk with only the relative strengths of the two components changing with position in the galaxy \citep{Nidever2014,Hayden2015}.

The origin of the $\alpha$-bimodality has been debated for decades.
The two-infall model of \citet{Chiappini1997} proposed that there were two episodes of gas infall that produced the two chemical populations.
In contrast, \citet{Schoenrich2009} stated that the chemically bimodal appearance comes about naturally due to the quick transition of stellar populations at early times from high-$\alpha$ to low-$\alpha$ and from radial mixing of stars at various radii in a disk containing a radial metallicity gradient.
\citet{Sharma2021} took this further by building an analytical chemodynamical model that includes radial mixing and was able to reproduce the observed Milky Way abundances with some fine-tuning of the parameters.

Alternatively, \citet{Clarke2019} showed that a disk with clumpy star formation could reproduce the $\alpha$-bimodality.  Early in the galaxy's evolution an instability forms clumps of gas with high star formation rate (SFR) that quickly self-enrich in $\alpha$-abundances and create the galaxy's high-$\alpha$ sequence.
At the same time, low-SFR star formation takes place throughout the galaxy and produces the low-$\alpha$ sequence.  
Another model put forward by \citet{Khoperskov2021} suggests that gas from early high-SFR outflows into the halo were re-accreted at later times reducing the metallicity of the gas and creating the low-$\alpha$ sequence.
Finally, there are some galaxy simulations that can reproduce the Milky Way's $\alpha$-bimodality by a gas-rich merger that brings in metal-poor gas around 8 Gyr ago and starts the formation of the low-$\alpha$ sequence \citep[e.g.,][]{Buck2020}.

Since we are now in a situation where multiple models can explain the Milky Way abundance data, 
we need a larger statistical sample than {\em one} to make further progress and rule-out models.
This can be achieved by measuring the $\alpha$-abundances for a Milky Way-like galaxy, such as M31.
Existing [$\alpha$/Fe] measurements of stars in M31 were primarily obtained with medium-resolution spectra ($R\sim 3000$ to 6000) using the DEIMOS multi-object spectrograph on the Keck~II telescope, and required either long exposure times ($\sim 6$~hours) to achieve S/N sufficient to measure a bulk [$\alpha$/Fe] abundance for individual stars with precision to $\sim 0.2$ to 0.4~dex \citep[e.g,][]{escala2019}, or co-addition of $\sim 5$ to 7 lower S/N stellar spectra to obtain mean abundance measurements of small groups of stars to a similar precision \citep{wojno2020}.
Most of the existing M31 abundances are in M31's stellar halo
\citep{escala2019,escala2020,gilbert2020,wojno2022}, tidal debris structures 
\citep{gilbert2019,escala2020dhs,escala2021},
and satellites \citep{kirby2020,wojno2020}.
Measurements of [$\alpha$/Fe] have been made in only one field in M31's stellar disk, at 26~kpc in projected radius; the 10 stars belonging to the disk component with [$\alpha$/Fe] abundances have an average [$\alpha$/Fe] of 0.6~dex with a standard deviation in [$\alpha$/Fe] abundances of 0.28~dex \citep{escala2020dhs}.

Even with many hours of integration time on the 10-m Keck telescopes, the M31 stellar $\alpha$-abundances are not precise enough to detect an $\alpha$-bimodality like the one seen in the Milky Way, for which a precision of $\sim$0.05 dex or better is needed. However, such a high abundance precision can be achieved by using JWST NIRSpec using the micro-shutter assembly (MSA) to obtain spectra of $\sim$100 stars in one pointing.  Although the spectral resolution is lower ($R$ $\sim$2700) than traditionally desired for chemical abundance work ($R$ $\geq$ 20,000), \citet{Ting2017} showed that precise abundances can be obtained with such resolutions as long as the signal-to-noise ratio (S/N) is high enough.  We describe here the initial results of our Cycle 1 JWST NIRspec program to measure chemical abundances in the M31 disk where we find {\em no} $\alpha$-bimodality.

\begin{figure}[t]
\begin{center}
    \includegraphics[height=4.7cm]{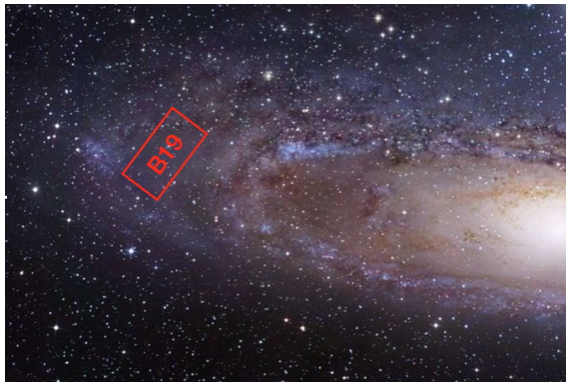}
    \includegraphics[height=5.1cm]{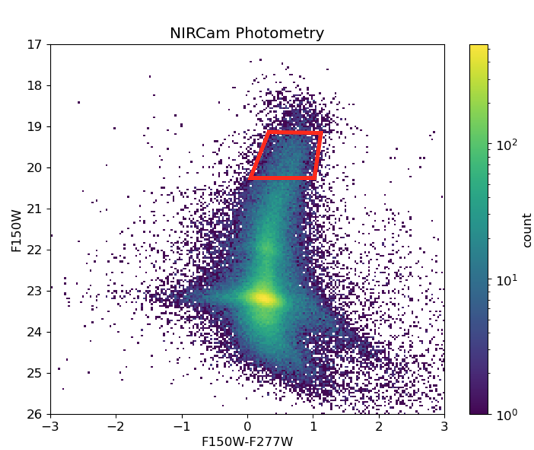}
    \caption{({\em Left}) Map of the M31 disk showing the location of our Brick 19 field (18 kpc projected from M31's center). ({\em Right}) Color-magnitude diagram of the NIRCam photometry showing our target selection of the top of the red giant branch.
    } \label{fig:mapcmd}
\end{center}
\end{figure}

\section{The JWST NIRSpec M31 Disk Program}
\noindent Our JWST Cycle 1 program (2609; PI: Nidever) observed one NIRSpec+MSA field in the southeastern M31 disk at a projected radius of 18 kpc.  We used the highest resolution grating G140H with the F100LP blocking filter that gave us a wavelength coverage of 9,000--18,000 \AA.  In addition, we observed Milky Way star cluster (M71 and IC166; with NGC~6791 to be observed in summer 2023) to help calibrate and validate the NIRSpec radial velocities (RVs) and chemical abundances.  These clusters have ground-based RV and chemical abundance results that can be compared to.

We determined that $\sim$100 stars and an abundance precision of $\sim$0.05 dex are needed to detect the $\alpha$-bimodality seen in the Milky Way.  Preliminary analysis with synthetic spectra indicated that this abundance precision could be achieved with the JWST NIRSpec resolution and wavelength coverage if a S/N of 70 was obtained.



\section{Data Reduction and Analysis}
\noindent The JWST CalSpec\footnote{\url{https://jwst-docs.stsci.edu/jwst-science-calibration-pipeline-overview}} pipeline automatically processes the 2-D NIRSpec images and performs spectral extraction as well as wavelength calibration.  It became immediately clear that there were several problems with the data processing, especially the spectral extraction which made the 1-D stellar spectra look like a sawtooth pattern.  Attempts to rerun the pipeline locally with different parameter setting proved ineffective.  Therefore, we developed a new Python software suite called \texttt{SPyderWebb}\footnote{\url{https://github.com/dnidever/spyderwebb}} that builds on the JWST CalSpec pipeline but greatly improves the data reduction.  In addition, it adds the capability to determine radial velocities with \texttt{Doppler} \citep{Doppler} as well as stellar parameters and chemical abundances with \texttt{FERRE} \citep{FERRE}.  The main components and improvements in \texttt{SPyderWebb} are:

\vspace{0.3cm}
\begin{itemize}
\item Better background subtraction.
\item Optimal extraction routines with empirical, non-parametric profiles.
\item Rejection of outlier pixels.
\item Slit-correction of wavelengths.
\item Radial velocity determination with \texttt{Doppler}.
\item Abundance determination using \texttt{FERRE}.
\end{itemize}

\vspace{0.3cm}
One revelation while working with the NIRSpec data was that due to the improved performance of the JWST telescope \citep{Rigby2023} the point spread function (PSF) on the NIRSpec detector is smaller than expected -- FWHM=0.9 pixels.  This means that the PSF is significantly undersampled compared to the nominally required Nyquist sampling of $\sim$2 pixels per PSF.  While this undersampling is not problematic for spectral extraction in the spatial dimension, it does mean that the 1-D spectra cannot be resampled onto a new wavelength scales because there is not enough information to do so.  We obtained six dithered exposures in our M31 field and were planning to combine these six ``visit'' spectra into one combined spectrum for each star.  Due to the undersampling, we instead had to ``forward model'' each visit spectrum using a model spectrum convolved with the correct PSF and sampled onto the observed wavelength scale.  Fortunately, this was straightforward to accomplish with \texttt{Doppler} and \texttt{FERRE}.

An unanticipated benefit of the PSF undersampling is that the spectral resolution is higher than expected.  For the G140H/F100LP setup, the predicted top resolution is $R$ $\sim$ 2,700. However, our analyses show that it is actually closer to $R$ $\sim$ 4,000--5,000.  This means that with proper analysis, higher precision is achievable with the RVs and chemical abundances.  In the future, it would be advisable for programs to spectrally dither by a $\sim$0.5 pixels to recover full sampling, as is done by APOGEE due to a slight undersampling in the blue part of the spectrum \citep{Nidever2015,Wilson2019}.

\begin{figure}[th]
    \includegraphics[width=0.33\textwidth]{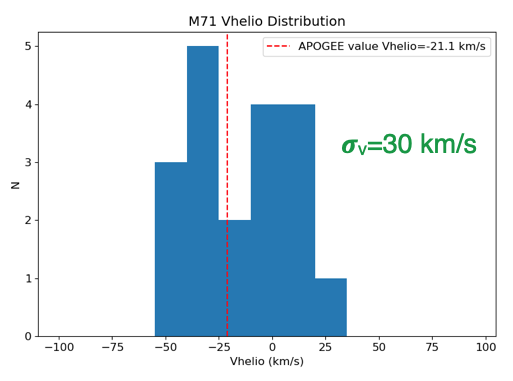}
    \includegraphics[width=0.34\textwidth]{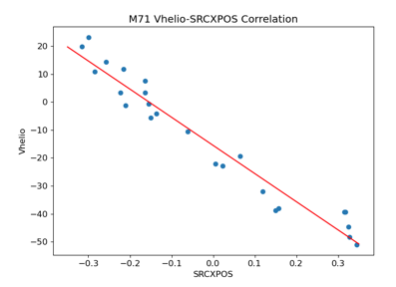}
    \includegraphics[width=0.31\textwidth]{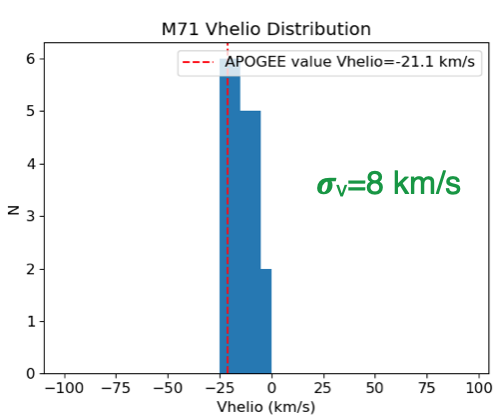}
    \caption{The histogram of radial velocities for our JWST NIRSpec M71 member stars. ({\em Left}) Original RV histogram with a scatter of $\sigma_V$=30 \kmse.
    ({\em Middle}) Correlation of the RVs with predicted shutter position. ({\em Right}) Corrected RV histogram with a significantly reduced scatter of $\sigma_V$=8 \kmse.
    } \label{fig:rvs}
\end{figure}

\vspace{0.1cm}
\noindent {\bf Radial Velocities:}
\noindent We used \texttt{Doppler} to determine precise radial velocities and estimates of the stellar parameters. \texttt{Doppler} forward-models a spectrum by convolving a model spectrum by the observed spectrum line-spread-function (LSF).  This allows us to handle the undersampled NIRSpec spectra, since the observed spectra do not need to be resampled.  \texttt{Doppler} uses a machine-learning \texttt{Cannon} \citep{Casey2016} model trained on a 3-D grid (\teffe, \logge, and [Fe/H]) of high-resolution synthetic spectra covering 3,000--18,000 \AA.  \texttt{Doppler} also has the ability to simultaneously fit multiple spectra of the same star with a single stellar model (``jointfit''), which we used on the multiple visit NIRSpec spectra.

Figure \autoref{fig:rvs} shows results for the NIRSpec M71 member radial velocities.  The left panel shows a histogram with a dispersion of 30 \kms which is substantially higher than the literature value \citep[2.2 \kmse;][]{Barth2020}.  The middle panel shows the M71 stellar RVs versus their predicted NIRSpec shutter position in the spectral dimension (\texttt{SRCXPOS}).  The strong correlation indicates that there is an RV offset due to the position of the stars in the ``slit''.  
The right panel shows the distribution of the slit-corrected RVs with a significantly smaller scatter of 8 \kmse.
The statistical uncertainties from \texttt{Doppler} are $\lesssim$1 \kmse, indicating that the scatter is still dominated by systematics.  We plan to investigate further ways to improve the RV and wavelength calibration (Tollerud et al., in prep.).

\begin{figure}[t]
    \includegraphics[width=\textwidth]{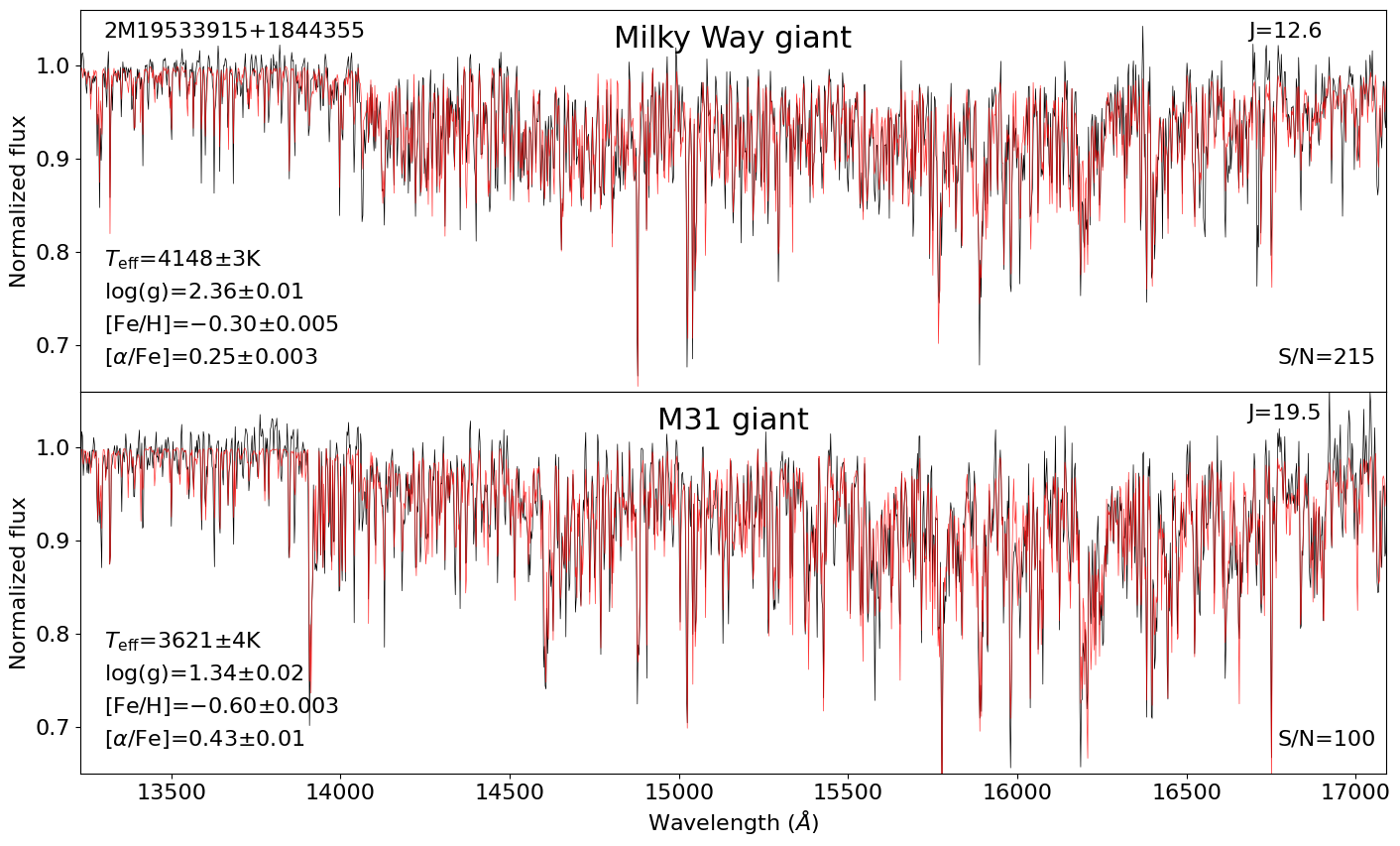}
    \caption{Example JWST NIRSpec spectra (black) and the best-fit \texttt{FERRE} synthetic spectrum (red). The top panel shows a Milky Way giant while the bottom panel shows a M31 giant seven magnitudes fainter but still with high S/N.} \label{fig:exampleferre}
\end{figure}

\vspace{0.5cm}
\noindent {\bf Abundances:}
\noindent We used \texttt{FERRE} to determine the stellar parameters, metallicity and $\alpha$-abundance.  The 4-D grid in \teffe, \logge, [M/H] and [$\alpha$/M] was generated for wavelengths 9,000--18,000 $\AA$ using the \texttt{Synspec} \citep{Hubeny2017} spectral synthesis package with Kurucz model atmospheres \citep{Kurucz2005}. \texttt{FERRE} was run in a mode where the synthetic spectrum is convolved with the correct LSF and resampled onto the wavelength scale of the observed spectrum.
Figure \autoref{fig:exampleferre} shows example JWST NIRSpec spectra (black) and their best-fit \texttt{FERRE} synthetic spectrum illustrating that reliable stellar parameters can be obtained.

\vspace{-0.05cm}
\section{M31 Abundances}
\begin{figure}[t]
\begin{center}
    \includegraphics[width=0.47\textwidth]{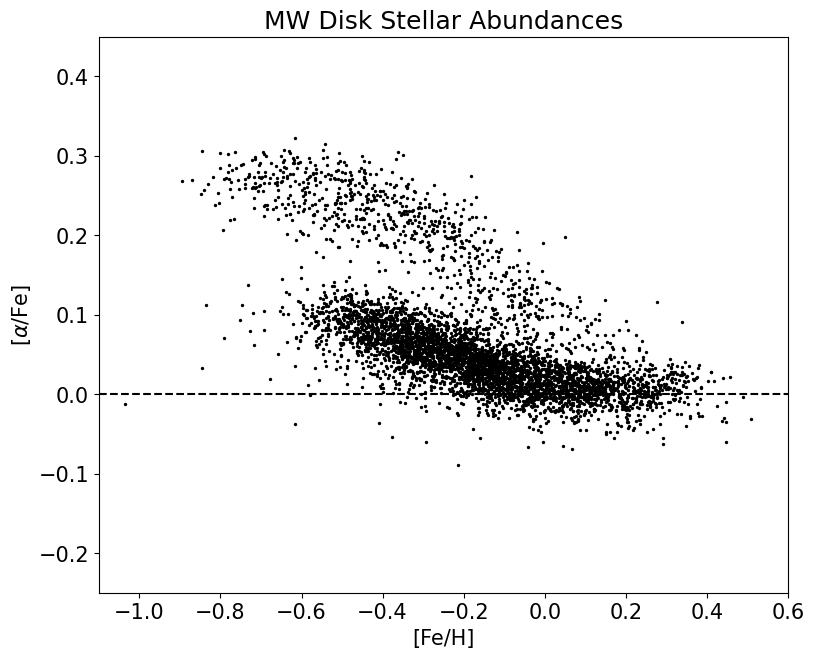}
    \includegraphics[width=0.47\textwidth]{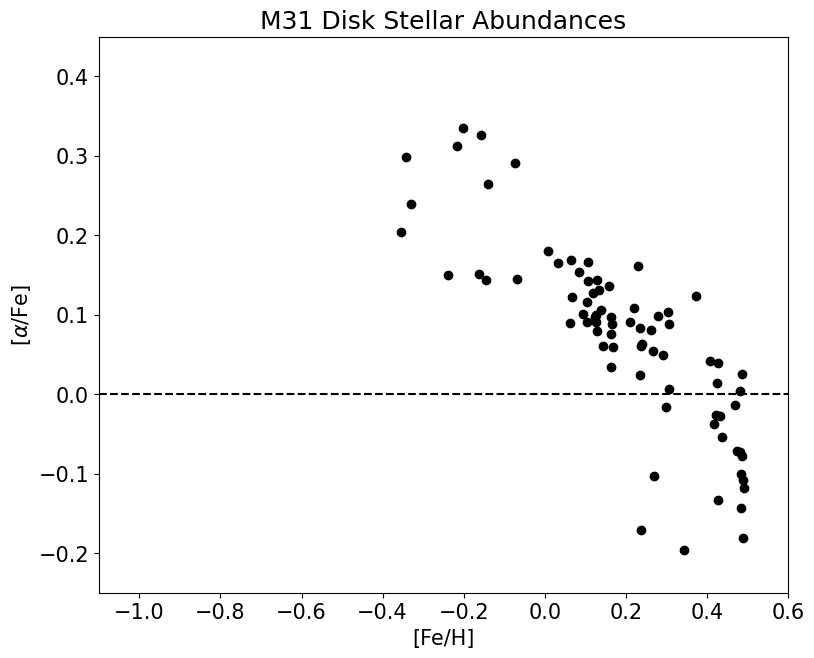}
    \caption{Comparison of the $\alpha$-abundances between the MW and M31. (Left) APOGEE red clump stars of the MW, and (right) M31 abundances from JWST NIRSpec.  While the MW show two sequences with an $\alpha$-bimodality, M31 only has one ``high-$\alpha$'' sequence with nothing like the MW's low-$\alpha$ sequence.} \label{fig:abundcomparison}
    \end{center}
\end{figure}

\noindent Figure \autoref{fig:abundcomparison} shows a comparison of the APOGEE red clump Milky Way $\alpha$-abundances \citep[from DR17;][]{Bovy2014} on the left and our NIRSpec M31 $\alpha$-abundances on the right.  While the MW has a prominent low-$\alpha$ population that extends from [Fe/H]=$-0.6$ to $+0.5$, that population does {\em not} exist in our M31 sample.  In fact, almost the entire stellar sample can be described by a single track similar to that of the MW's high-$\alpha$ population which is associated with old stars in the thick disk and a high SFR.  Therefore, there is {\em no} $\alpha$-abundance bimodality in the M31 disk.

While this finding is somewhat surprising, it is consistent with other recent results.  \citet{Dalcanton2023} used the PHAT photometric survey to find that the M31 southeastern disk is dominated by a thick component with scale-height of 0.77 kpc, which is similar to the Milky Way's thick disk \citep[$\sim$1 kpc;][]{BlandHawthorn2016}.
In addition, \citet{Dorman2015} found a higher velocity dispersion in the M31 disk of $\sim$60 \kms which is multiple times higher than the velocity dispersion in the MW's older components.
This difference between the MW and M31's disks is likely due to M31's higher merger and accretion rate.  Evidence of this includes the
Giant Southern Stream \citep[GSS;][]{Ibata2001} and the metal-rich inner halo that was likely produced by a recent merger only $\sim$2 Gyr ago \citep{DSouzaBell2018,Hammer2018}.
Therefore, it understandable that the M31 disk is dominated by a thick component.

What does this mean for the $\alpha$-bimodality models previously mentioned?  While mergers and interactions will ``puff-up'' or dynamically heat a stellar disk, this will not change the chemistry of the already-existing stars or change internal processes such as basic chemical evolution.  However, the mergers might have changed certain conditions required by the models.  For example, the \citet{Schoenrich2009} and \citet{Sharma2021} models require a radial metallicity gradient.  While the MW has a fairly strong gradient \citep[$-0.068$ dex/kpc;][]{Donor2020}, M31's is factor of 3.4$\times$ lower \citep[$-$0.02 dex/kpc;][]{Gregersen2015,Pena2019,Escala2023} which has been attributed to mergers and interactions.
These models also require radial migration to create the low-$\alpha$ sequence that is very extended in metallicity, but radial migration is generally less efficient in a dynamically hot disk \citep{Sellwood2002,Solway2012}.
On the other hand, the clumpy star formation model requires conditions that allow for the clump instability.  It's quite possible that these conditions were not met in the M31 disk due to the active merger history. Therefore, all of the models will need to be tested in the conditions of M31 to ascertain if they can explain the existence of a bimodality in the MW but the lack of one in M31.

\section{Conclusions}
\noindent We present our initial results from our JWST NIRSpec M31 disk project.  We observed one field and obtained high-S/N spectra of 103 RGB stars.  For these stars, we were able to determine stellar parameters and precise $\alpha$-abundances.  While the Milky Way has two $\alpha$-abundance sequences (low-$\alpha$ and high-$\alpha$) with an $\alpha$-bimodality at intermediate metallicities, our M31 results show nothing like the MW's low-$\alpha$ sequence or the $\alpha$-bimodality.  In fact, the M31 abundances can be explained by a single high-$\alpha$ population formed with a high star formation rate.  These results are consistent with other recent findings in the literature that conclude that the M31 disk is dominated by a thick, high velocity dispersion stellar population.  The difference between the MW and M31 disks are likely driven by the higher merger and accretion rate of M31.  These contrasts suggest that the dominant processes at work in forming the chemistry and structure in M31's disk were somewhat different than in the Milky Way.

While calibration work is still needed to realize its full potential, we believe that the capabilities of JWST with NIRSpec/MSA for stellar spectroscopy work will produce precise radial velocities and chemical abundances for many galaxies in the Local Group and beyond and will produce important advances in our understanding of galaxy formation and evolution.

\section*{Acknowledgements}
\noindent We thank the conference organizers for putting together such an interesting conference.  Our greatest thanks goes to the JWST, NIRSpec and STScI teams for constructing such an amazing telescope and instrument.


\end{document}